\documentstyle[12pt,epsf,aps]{revtex}

\normalsize

\newcommand{\Qa}{Q_{\alpha}}
\newcommand{\dlt}{\Delta_n/\Delta_p}

\begin{document}

\title{\large {\bf The structure of superheavy elements 
      newly discovered in the reaction 
  of $^{86}$Kr with $^{208}$Pb 
} }
\author{J. Meng$^{a,b,c}$ and N.Takigawa$^{b}$  \\
     $^{a}$Department of Technical Physics, Peking University,\\
               Beijing 100871, P.R. China \\
      $^{b}$Department of Physics, Tohoku University\\
               Sendai 980-8578, Japan \\
      $^{c}$Center of Theoretical Nuclear Physics, National Laboratory of \\
       Heavy Ion Accelerator, Lanzhou 730000, China\\
         e-mail: meng@ihipms.ihip.pku.edu.cn}
\date{\today}
\maketitle

\vspace{1.5cm}
\begin{abstract}
\indent
The structure of superheavy elements newly discovered 
in the $^{208}$Pb($^{86}$Kr,n) reaction at Berkeley 
is systematically studied in the Relativistic Mean Field (RMF) approach. 
It is shown that various usually employed 
RMF forces, which give fair description of normal stable nuclei, 
give quite different predictions for superheavy elements. 
Among the effective forces we tested, TM1 is found to be the 
good candidate to describe superheavy elements. 
The binding energies 
of the $^{293}$118 nucleus and its $\alpha-$decay daughter nuclei 
obtained using TM1 agree with
those of FRDM within $2$ MeV.
Similar conclusion that TM1 is the good interaction is 
also drawn from the calculated binding energies 
for Pb isotopes with the 
Relativistic Continuum Hartree Bogoliubov (RCHB) theory.
Using the pairing gaps obtained from RCHB, 
RMF calculations with pairing and deformation 
are carried out for the structure of superheavy elements.
The binding energy, shape, single particle levels, 
and the Q values of the $\alpha-$decay $Q_{\alpha}$ 
are discussed, and it is shown that 
both pairing correlation and deformation 
are essential to properly understand the structure 
of superheavy elements. 
A good agreement is obtained 
with experimental data on $Q_{\alpha}$. 
\end{abstract}




\section{Introduction}

Following the discovery of 
$\alpha-$decay isotopes of 
Elements $Z = 110, 111$, and $112$
at GSI 
\cite{Hof.95,Hof.95a,Hof.96}, 
an isotope of the Element 118, $^{293}$118, and 
several its $\alpha-$decay daughter nuclei 
were announced to have been discovered
at Berkeley Lab's 88-Inch Cyclotron 
with the newly constructed Berkeley Gas-filled Separator
by bombarding 
lead target with an intense beam of 
krypton ions 
of 449 MeV \cite{Nino99}. 
The sequence of decay events 
is consistent with the long-standing theoretical prediction that there exists
an ``island of stability''
around 114 protons
and 184 neutrons and 
activates once again the 
study of superheavy elements. 

The study of 
superheavy elements has been a 
hot topic for the last two decades.  
Recent works on the collisions, 
structure and stability of Heavy and Superheavy Elements
can be found 
in Refs. \cite{IMRS.96,WGF.96,LSRG.96,CDHMN.96,RBBS.97,AASV.98,BRRMG.98}.
In a recent paper, 
Smolanczuk
claimed that the reaction $^{208}Pb(^{86}Kr,n)$  should have a
particularly favorable production rate \cite{smo99}. 
This motivated the experiment at Berkeley. According to 
the authors, the synthesized superheavy element $^{293}$118 decays 
by emitting an alpha 
particle within less than a millisecond, 
leaving behind the isotope of element 116 
with mass number 289. 
This daughter nucleus
is also radioactive, 
alpha-decaying to an isotope of element 114. 
The chain of successive alpha decays continues until 
element 106. 

Smolanczuk discussed also the properties of superheavy elements 
in this mass region under the constraint of a spherical shape
based on a macroscopic-microscopic approach
\cite{smo97}. In contrast to his approach, 
here we study the structure of superheavy element 
$^{293}$118
and of the daughter nuclei in the sequence of $\alpha$ decays
in the Relativistic Mean Field (RMF) theory. 
The effects of deformation and pairing 
correlation will be taken into account. 
The pairing gaps for deformed RMF 
calculations are taken from 
the Relativistic Continuum
Hartree Bogoliubov (RCHB) theory \cite{ME.98}, which is an extension of the
Relativistic Mean Field and the Bogoliubov transformation
in the coordinate representation \cite{MR.96}.
As the spin-orbit
splitting which governs the shell structure 
and magic number 
is naturally obtained in the RMF theory, we 
expect that the structure of superheavy 
elements 
can be understood properly once the deformation and 
pairing correlation 
are taken into account.
We investigate 
the binding energy, 
deformation, the $Q$-values of the alpha decay, 
the effect of pairing correlation,
shell structure, and 
the structure of single particle 
levels for protons and neutrons.

The paper is organized as follows. 
In sect.II, we present the results of RMF
calculations without pairing correlation 
for several standard 
forces, which give fair description of normal
stable nuclei. We thus discuss the appropriate force to describe 
superheavy elements.
In sect.III the RCHB theory  
is used to investigate the pairing correlation in these superheavy 
elements. The RCHB provides not only a unified
description of mean field and pairing correlation
but also a proper description for the continuum and the
coupling between the bound state and the continuum \cite{ME.98}.
We then perform in sect.IV the study by a deformed RMF+BCS approach 
using the pairing gaps supplied by RCHB. 
We summarize the paper in sect.V.


\section{Examination of various RMF parameter sets}

There are many parameter sets for RMF calculations, which 
provide nearly equal quality of description for stable nuclei. 
Therefore, we wish to find at first which effective force in RMF is more 
suitable to describe superheavy elements. As claimed in 
Ref.\cite{BRRMG.98} the results are strongly interaction 
dependent. For this purpose we perform 
RMF calulations 
that include deformation but ignore pairing correlation 
with different effective forces. 
The details of the method can be found in Ref.\cite{GRT.90}.
  
Table \ref{DRMFE} compares
the binding energies $E$ of superheavy element $^{293}118$ 
and its $\alpha$ decay daughter nuclei calculated
with effective forces TM1 \cite{Toki}, NL1, NL3 and NLSH. 
For comparison, 
the results of the phenomenological FRDM calculations 
are given in the last column\cite{Mol.95}.

The results of TM1 are nearly the same as those of FRDM 
for $^{265}104$, $^{269}106$, $^{273}108$, 
$^{277}110$,  and $^{281}112$.  
They are within 1 MeV from each other.
The difference 
between TM1 and FRDM results gets larger for $^{285}114$,
$^{289}116$ and $^{293}118$, but is still 
smaller than 3 MeV. Though there are differences of several MeV, 
NL1 and NLSH give similar results as TM1. 
The NL3 parameter set, on the other hand,
gives a difference of about 50 MeV from the other calculations. 

One important difference between the RMF calculations with TM1 and 
FRDM is that the additional gain of the binding energy when one moves from 
$^{281}$112 to $^{285}$114 is much less in the RMF calculations. 
In other words, Z=114 has a weaker meaning as a magic number in the RMF 
calculations. Though, strictly speaking, it may not be adequate, 
let us call this effect as the change of the shell structure 
or of the magic number property at Z=112. 
As we see shortly, a similar effect appears in the Z dependence of the 
nuclear shape. This effect eventually plays
an important role in reproducing the qualitative trend of the experimental 
data on the atomic number dependence of $Q_{\alpha}$.


\begin{table}
\caption{The binding energy $E$ of 
the superheavy element $^{293}$118 and 
of its $\alpha$ decay daughter nuclei 
calculated in the RMF theory with different effective 
interactions TM1, NL1, NL3, and NLSH.
The prediction of FRDM is also given in the last column.
}
\begin{center}
\label{DRMFE}
\begin{tabular}{ccccccccccc}
$^A    Z   _N$    && TM1    & NL1    & NL3    & NLSH   & FRDM   & \\
\hline \hline
$^{265}104_{161}$ && 1949.2 & 1953.0 & 1912.5 & 1950.5 & 1950.0 &  \\
\hline
$^{269}106_{163}$ && 1970.2 & 1974.1 & 1932.0 & 1977.5 & 1970.5 &  \\
\hline
$^{273}108_{165}$ && 1990.2 & 1992.8 & 1950.9 & 1997.9 & 1989.4 &  \\
\hline
$^{277}110_{167}$ && 2007.9 & 2010.5 & 1968.2 & 2016.1 & 2007.0 &  \\
\hline
$^{281}112_{169}$ && 2026.2 & 2029.8 & 1986.2 & 2034.7 & 2025.2 &  \\
\hline
$^{285}114_{171}$ && 2041.3 & 2051.8 & 2003.0 & 2048.7 & 2044.1 &  \\
\hline
$^{289}116_{173}$ && 2058.2 & 2068.7 & 2014.2 & 2065.6 & 2061.1 &  \\
\hline
$^{293}118_{175}$ && 2074.7 & 2088.9 & 2028.7 & 2080.5 & 2077.2 &  \\
\end{tabular}
\end{center}
\end{table}

Similar trend concerning the mutual comparison of different 
forces appears also in Table \ref{RMFQE},
where the Q-values of $\alpha$ decay sequence 
$Q_{\alpha}=E(^4He)+E(Z-2,N-2)-E(Z,N)$ (MeV) 
are shown. The Q-values given by TM1 and FRDM are quite similar 
except for $^{285}114$, where the difference is 3.8 MeV.
This large difference is connected with the 
change of the shell structure mentioned above.

\begin{table}
\caption{The Q-values of $\alpha$ decay
for superheavy element $^{293}118$ and its $\alpha$ decay daughter nuclei.
We used the experimental value
for the binding energy of the $\alpha$ particle.
}
\begin{center}
\label{RMFQE}
\begin{tabular}{ccccccccccc}
$^A    Z   _N$    && $TM1$ & $NL1$ & $NL3$ & $NLSH$ &  $FRDM$  & \\
\hline \hline
$^{269}106_{163}$ && 8.3   & 7.2   & 8.8   & 1.3    &  7.8   &  \\
\hline
$^{273}108_{165}$ && 8.3   & 9.6   & 9.4   & 7.9    &  9.4   &  \\
\hline
$^{277}110_{167}$ && 10.6  & 10.6  & 11.0  & 10.1   &  10.7  &  \\
\hline
$^{281}112_{169}$ && 10.0  & 9.0   & 10.3  & 9.7    &  10.1  &  \\
\hline
$^{285}114_{171}$ && 13.2  & 6.3   & 11.5  & 14.3   &  9.4   &  \\
\hline
$^{289}116_{173}$ && 11.4  & 11.4  & 17.1  & 11.4   &  11.3  &  \\
\hline
$^{293}118_{175}$ && 11.8  & 8.1   & 13.8  & 13.4   &  12.2  &  \\
\end{tabular}
\end{center}
\end{table}

Table \ref{DRMFBeta} shows the corresponding 
deformation parameter $\beta$ 
in the ground state.
TM1 predicts a stable prolate deformation $\beta \sim 0.2$ for all the 
nuclei listed in the table, taking the minimum at 
$^{281}112$.
NL3 and NLSH give similar results as TM1, but the 
minimum deformation is shifted to $^{285}114$ for NL3. 
The NL1 predicts a spherical shape for 
$^{293}118$, while FRDM almost spherical shape for 
$^{277}110$, $^{281}112$, 
$^{285}114$, $^{289}116$ and $^{293}118$. 
The shift of the atomic number, where the deformation becomes minimum, 
from Z=114 to 112 is what we already mentioned 
as an evidence of the change of the shell structure.

\begin{table}
\caption{The deformation $\beta$ 
of the superheavy element 
$^{293}$118 and its $\alpha$ decay daughter nuclei
calculated with different effective
interactions TM1, NL1, NL3, and NLSH. 
The prediction of FRDM is also given in the last column.
}
\begin{center}
\label{DRMFBeta}
\begin{tabular}{ccccccccccc}
$^A    Z   _N$    && $TM1$  & $NL1$  & $NL3$  & $NLSH$ &  $FRDM$  & \\
\hline \hline
$^{265}104_{161}$ && 0.2656 & 0.2687 & 0.2685 & 0.2513 &  0.222 &  \\
\hline
$^{269}106_{163}$ && 0.2059 & 0.2576 & 0.2479 & 0.2073 &  0.221 &  \\
\hline
$^{273}108_{165}$ && 0.2021 & 0.2438 & 0.2032 & 0.2068 &  0.173 &  \\
\hline
$^{277}110_{167}$ && 0.1857 & 0.2279 & 0.1865 & 0.1914 &  0.089 &  \\
\hline
$^{281}112_{169}$ && 0.1681 & 0.2072 & 0.1683 & 0.1741 &  -.096 &  \\
\hline
$^{285}114_{171}$ && 0.2118 & 0.1593 & 0.1606 & 0.2316 &  0.080 &  \\
\hline
$^{289}116_{173}$ && 0.2373 & 0.1548 & 0.2163 & 0.2566 &  0.081 &  \\
\hline
$^{293}118_{175}$ && 0.2340 & 0.0600 & 0.2402 & 0.2946 &  0.080 &  \\
\end{tabular}
\end{center}
\end{table}

Table \ref{RMFRc} compares the corresponding charge-radii $R_c$. 
In contrast to the big difference seen in the binding energy, the 
charge-radii $R_c$ for different forces lie within $1\%$ 
from each other.

\begin{table}
\caption{Comparison of the charge-radii $R_c$ of superheavy element 
$^{293}118$ and its $\alpha$ decay daughter nuclei
calculated with different parameter sets.
}
\begin{center}
\label{RMFRc}
\begin{tabular}{||cc|c|c|c|c||}
$^A    Z   _N$    && $TM1$ & $NL1$ & $NL3$ & $NLSH$  \\
\hline \hline
$^{269}106_{163}$ && 6.146 & 6.153  & 6.177 & 6.104  \\
\hline
$^{273}108_{165}$ && 6.174 & 6.176 & 6.197 & 6.134  \\
\hline
$^{277}110_{167}$ && 6.197 & 6.203 & 6.220 & 6.158  \\
\hline
$^{281}112_{169}$ && 6.226 & 6.228 & 6.247 & 6.188  \\
\hline
$^{285}114_{171}$ && 6.282 & 6.248 & 6.275 & 6.250  \\
\hline
$^{289}116_{173}$ && 6.332 & 6.272 & 6.340 & 6.303  \\
\hline
$^{293}118_{175}$ && 6.360 & 6.276 & 6.392 & 6.355  \\
\end{tabular}
\end{center}
\end{table}


\section{Pairing correlation in superheavy elements: description by RCHB}

In this section we study the effects of 
pairing correlation in superheavy element 
$^{293}$118
and its $\alpha$ decay daughter nuclei 
by using the self-consistent and fully 
microscopic RCHB theory\cite{ME.98} 
under the constraint of a spherical shape. 
With the pairing gap obtained from RCHB, 
a self-consistent and more complete RMF calculation 
with both pairing correlation and deformation will be
carried out in the next section.

Before applying the RCHB theory to newly discovered superheavy elements, 
we examine once again which effective force is the most suitable 
to describe superheavy elements. 
For this purpose, we use lead isotopes as test cases. 

\begin{table}
\caption{Comparison of the binding energies $E$ of Pb isotopes 
calculated in RCHB theory 
with 4 different parameter sets with experimental data. 
The last four columns are 
the root mean square neutron, proton, matter and charge radii calculated 
with the TM1 parameter set.}
\begin{center}
\label{RCHBPb}
\begin{tabular}{||cc|c|c|c|c|c|c|c|c|c||}
A(Pb)&& Exp.[MeV] & TM1   & NL1     & NL3      & NLSH  &  $R_N$  &    $R_P$   &   $R_M$ &    $R_C$ \\
\hline \hline
202  && 1592.20 & 1592.91 & 1596.60 & 1565.43 & 1596.00 & 5.629 &   5.420  & 5.545  &  5.479 \\
\hline
204 && 1607.52  & 1609.18 & 1611.35 & 1580.36 & 1611.35 & 5.656 &   5.429  & 5.566  &  5.487 \\
\hline
206 && 1623.40  & 1623.78 & 1625.73 & 1594.77 & 1626.11 & 5.683 &   5.437  & 5.586  &  5.495 \\
\hline
208 && 1636.45  & 1637.76 & 1639.72 & 1608.56 & 1639.99 & 5.713 &   5.447  & 5.609  & 5.505  \\
\hline
210 && 1645.57  & 1646.92 & 1646.78 & 1618.04 & 1648.34 & 5.743 &   5.467  & 5.636  & 5.525  \\
\hline
212 && 1654.52  & 1655.73 & 1653.57 & 1626.97 & 1656.43 & 5.772 &   5.486  & 5.663  & 5.544 \\
\end{tabular}
\end{center}
\end{table}

The binding energies of six Pb isotopes 
calculated by RCHB with 4 different effective forces 
are compared with experimental data in Table \ref{RCHBPb} .
Although all the calculations except for NL3 
well reproduce the experimental binding energies of the Pb isotopes,
TM1 gives the best reproduction of the data.
Therefore we expect that the RMF calculations with TM1 and 
pairing correlation  
will give a satisfactory description of superheavy elements.
The $rms$ radii for neutron $R_N$,
proton $R_P$, matter $R_M$, and charge radii $R_C$
calculated by RMF with TM1 are
given in the last four columns in Table \ref{RCHBPb}.

\begin{table}
\caption{The binding energy $E$, one neutron separation energy $S_n$,
the Q value for the $\alpha$ decay $\Qa$, matter and charge 
$rms$ radii $R_m$ and $R_c$, neutron and proton pairing gaps
in RCHB with TM1 for the superheavy element
$^{293}$118 and its $\alpha$ decay daughter nuclei. 
The results for $^{86}$Kr, $^{208}$Pb and $^{294}$118 are 
also given.  
}
\begin{center}
\label{RCHBTM1}
\begin{tabular}{||cc|c|c|c|c|c|c||}
$^A    Z   _N$    &&   $E$  & $S_n$ & $\Qa$ & $R_m$ & $R_c$ & $\dlt$       \\
\hline \hline
$^{ 86} Kr_{ 50}$ &&  750.2 &       &       & 4.221 &  4.182 & -0.010/--1.304 \\
\hline
$^{208} Pb_{126}$ && 1637.6 &       &       & 5.649 &  5.541 & -0.000/-0.000 \\
\hline
$^{265}104_{161}$ && 1944.4 &       &       & 6.175 &  6.099 & -0.622/-1.173 \\
\hline
$^{269}106_{163}$ && 1965.6 & 4.9   & 7.1   & 6.202 &  6.131 & -0.421/-1.146 \\
\hline
$^{273}108_{165}$ && 1986.8 & 5.9   & 7.1   & 6.228 &  6.160 & -0.283/-1.133 \\
\hline
$^{277}110_{167}$ && 2007.2 & 5.5   & 7.9   & 6.255 &  6.189 & -0.380/-1.092 \\
\hline
$^{281}112_{169}$ && 2027.1 & 6.3   & 8.4   & 6.281 &  6.218 & -0.338/-1.030 \\
\hline
$^{285}114_{171}$ && 2046.8 & 6.6   & 8.6   & 6.308 &  6.247 & -0.030/-0.948 \\
\hline
$^{289}116_{173}$ && 2065.2 & 5.8   & 9.9   & 6.333 &  6.273 & -0.013/-0.841 \\
\hline
$^{293}118_{175}$ && 2082.1 & 5.9   & 11.5  & 6.356 &  6.296 & -0.315/-0.696 \\
\hline
$^{294}118_{176}$ && 2088.8 & 6.7   &       & 6.364 &  6.299 & -0.442/-0.702 \\
\end{tabular}
\end{center}
\end{table}

We have then calculated 
the binding energy $E$, one neutron separation energy $S_n$,
the Q value for the $\alpha$ decay $\Qa$, matter and charge $rms$ radii 
$R_m$ and $R_c$, neutron and proton pairing gaps
for superheavy elements in the RCHB with TM1. The results are 
shown in Table \ref{RCHBTM1}. 
The matter $rms$ radius $R_m$ is larger than
the charge $rms$ radius $R_c$ for all nuclei due to the neutron excess. 
The Q value of the $\alpha$ decay increases monotonically
with $Z$. The proton pairing gap parameter is around 1 MeV, while
the neutron pairing gap parameter is relatively small due to
blocking effects.
The calculation for $^{ 86} Kr$, $^{208} Pb$ and
$^{294}118$ are also given for reference 
to understand the fusion barrier to synthesize 
the element $^{294}118$.

\begin{figure}
{\centerline{
     {\epsfxsize 15 cm \epsffile{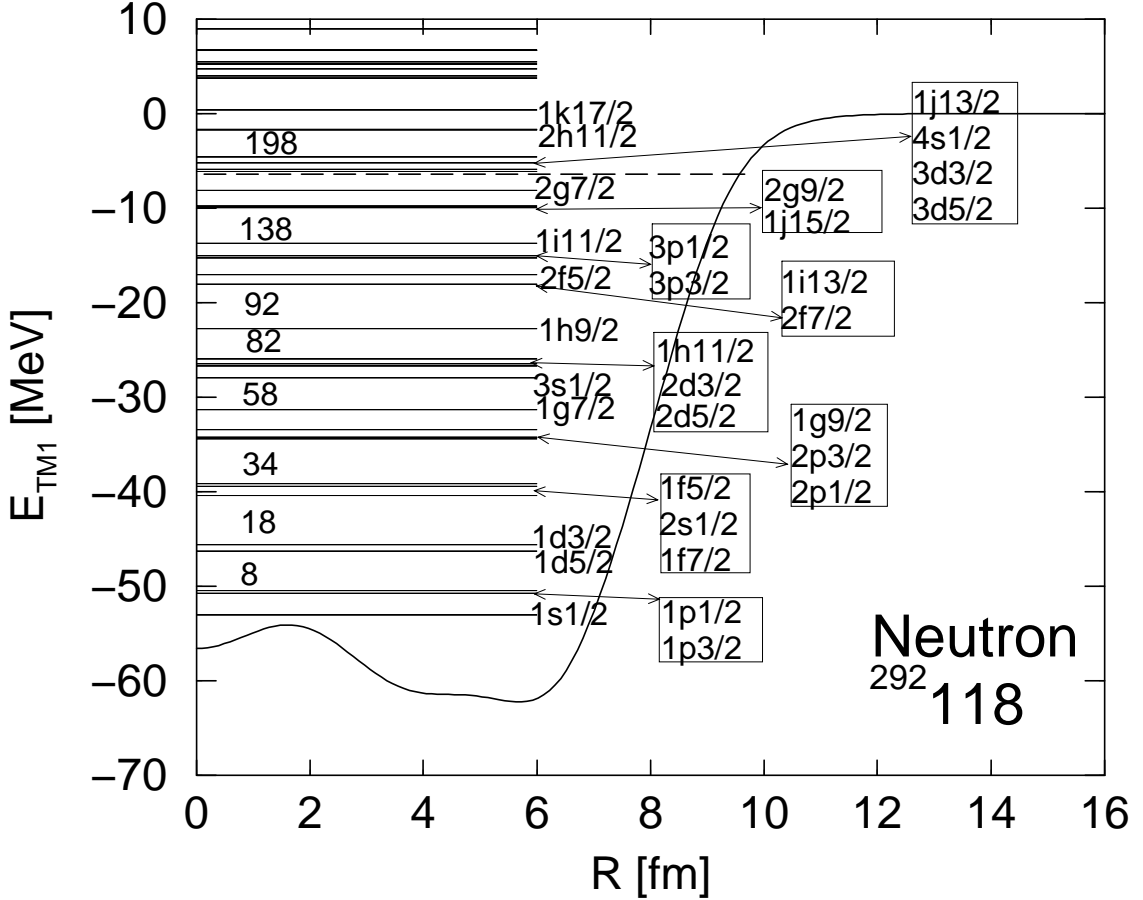}}}}
\vspace{8mm}
\caption{The single particle levels
in the canonical basis for neutrons in
$^{292}$118 calculated by RCHB with TM1.
The neutron potential $V_V( r ) + V_S( r )$
is represented by the solid line and the Fermi level
by a dashed-line.
}
\label{RCHBnl}
\end{figure}

\begin{figure}
{\centerline{
     {\epsfxsize 15 cm \epsffile{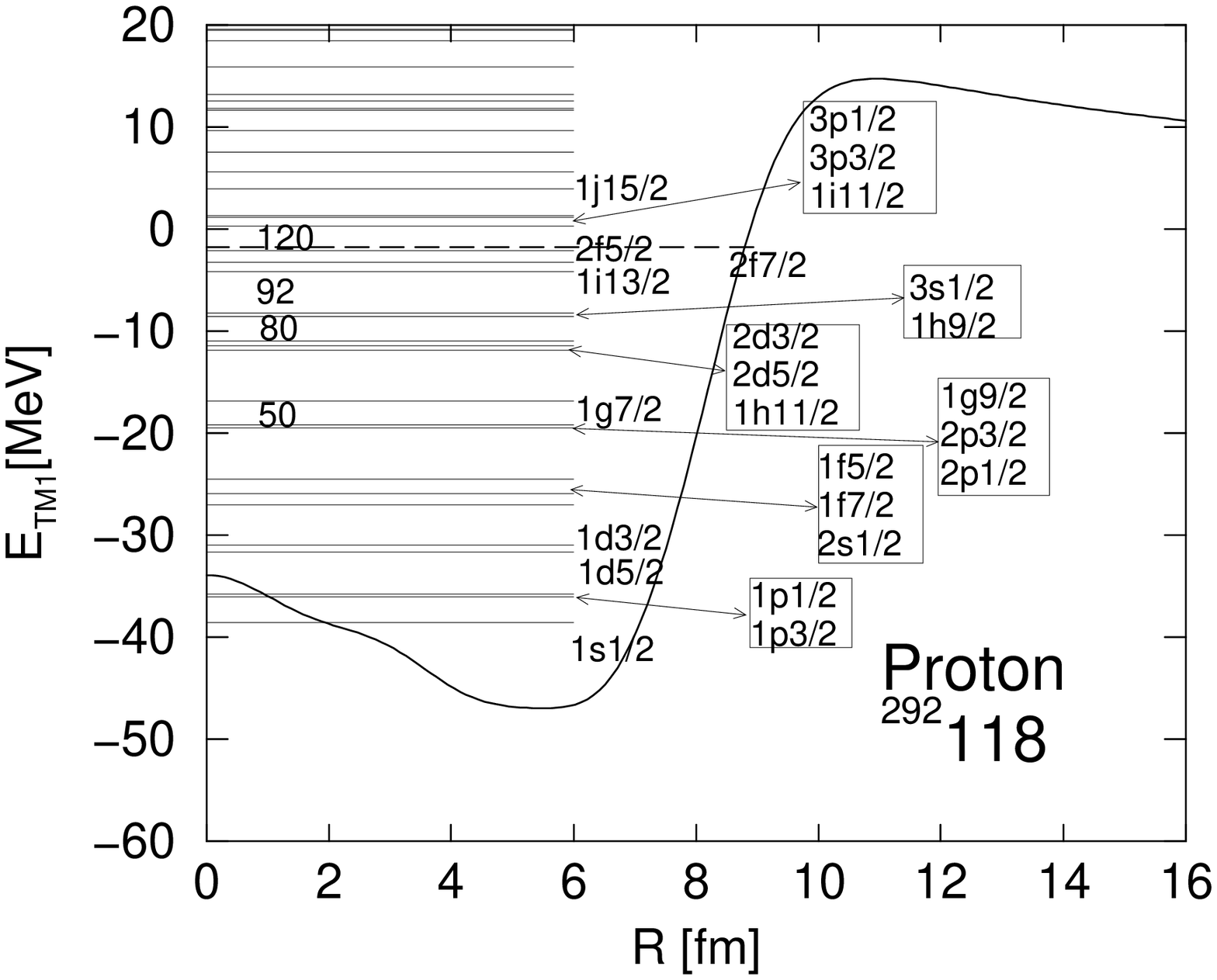}}}}
\vspace{8mm}
\caption{The same as Fig.1, but for protons.}
\label{RCHBpl}
\end{figure}

In Figs.\ref{RCHBnl} and \ref{RCHBpl}, the single particle levels
in the canonical basis for neutrons and protons in $^{292}$118  
are given, respectively. 
In order to avoid the 
irregularity due to the blocking effect, we give the 
single particle levels in $^{292}$118 instead of $^{293}118$.
The Fermi surface for 
neutrons and protons is given in each figure by the dashed line.
The potential is the sum of the vector and scalar 
potentials.  
Fig.\ref{RCHBnl} indicates that, after the sub-closed shell at $N = 164$, 
the next closed or sub-closed shells occur at $N = 198$ and 
$N = 210$. For the proton case, closed or sub-closed shells
occur at $Z = 106$, $Z = 114$ and $Z = 120$. 

The Fermi level for protons in $^{293}118$ is at 
$\lambda = -1.916$ MeV, while that for neutrons at
$\lambda = -6.304$ MeV. Although the Fermi level 
for protons is very close to the continuum, 
the wave functions of all the protons are well localized 
in a small region because of the 
Coulomb barrier.

\begin{figure}
{\centerline{
     {\epsfxsize 15 cm \epsffile{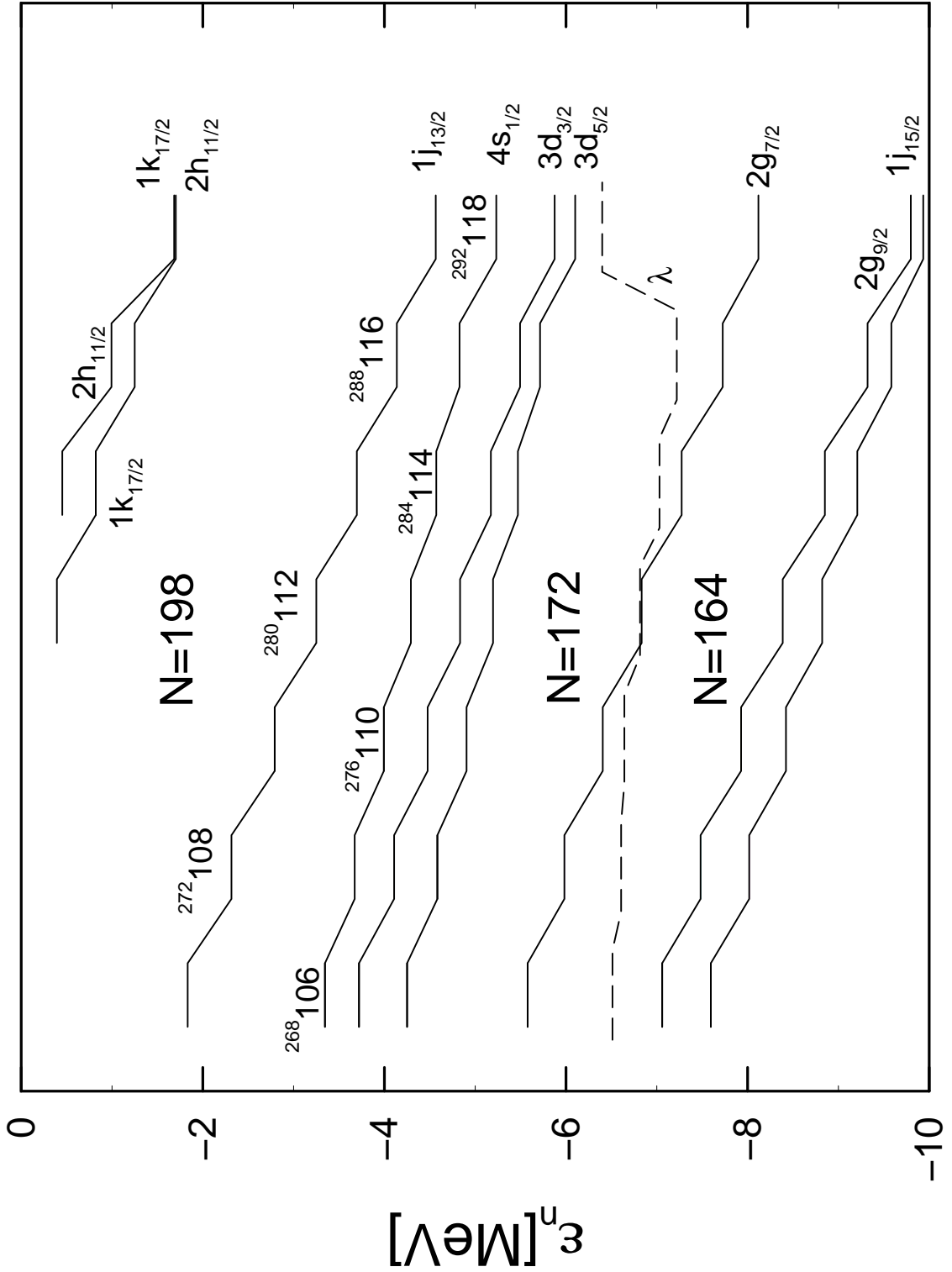}}}}
\vspace{8mm}
\caption{The single particle levels for neutrons near the Fermi surface
calculated in the RCHB with TM1. They are shown  
for the neighboring even-even nuclei to the superheavy elements in 
the $\alpha-$decay chain from $^{293}$118.}
\label{RCHBnal}
\end{figure}

\begin{figure}
{\centerline{
     {\epsfxsize 15 cm \epsffile{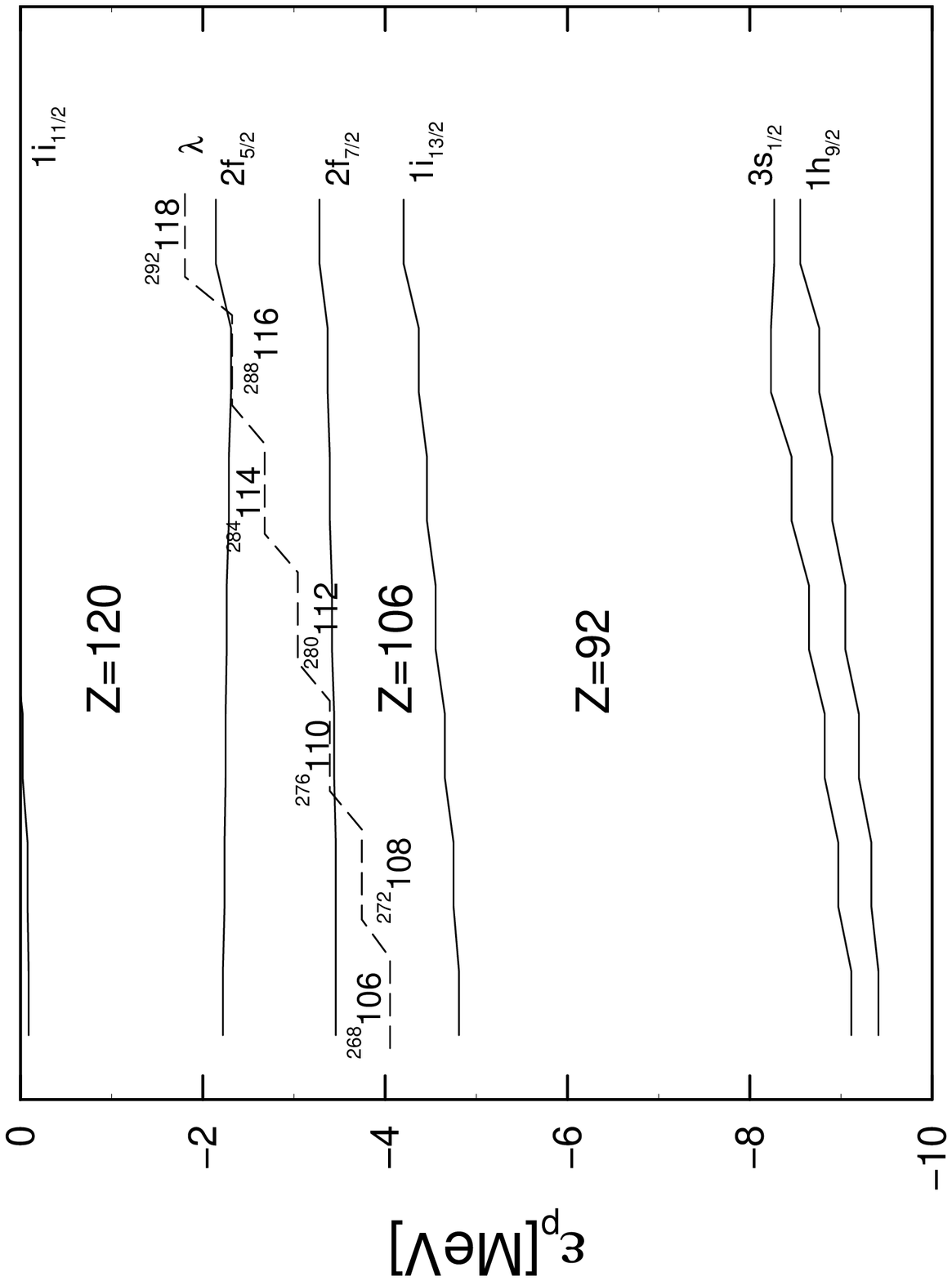}}}}
\vspace{8mm}
\caption{The same as Fig.3, but for protons.}
\label{RCHBpal}
\end{figure}

Figs.\ref{RCHBnal} and \ref{RCHBpal} show the change of 
the single particle neutron and proton levels near the Fermi surface
along the $\alpha-$decay chain from $^{293}$118. 
Similarly to Figs.\ref{RCHBnl} and \ref{RCHBpl}, we give the
single particle levels for the neighboring 
even-even nuclei in order to avoid the 
irregularity due to the blocking effect. Adding an $\alpha$ 
particle always raises the proton single particle levels 
and lowers the neutron single particle levels. 
There are distinct gaps of about 2 MeV at 
$N=164$ and $172$ and of about 3 MeV at
$N=198$ for neutrons. 

\begin{figure}
{\centerline{
     {\epsfxsize 15 cm \epsffile{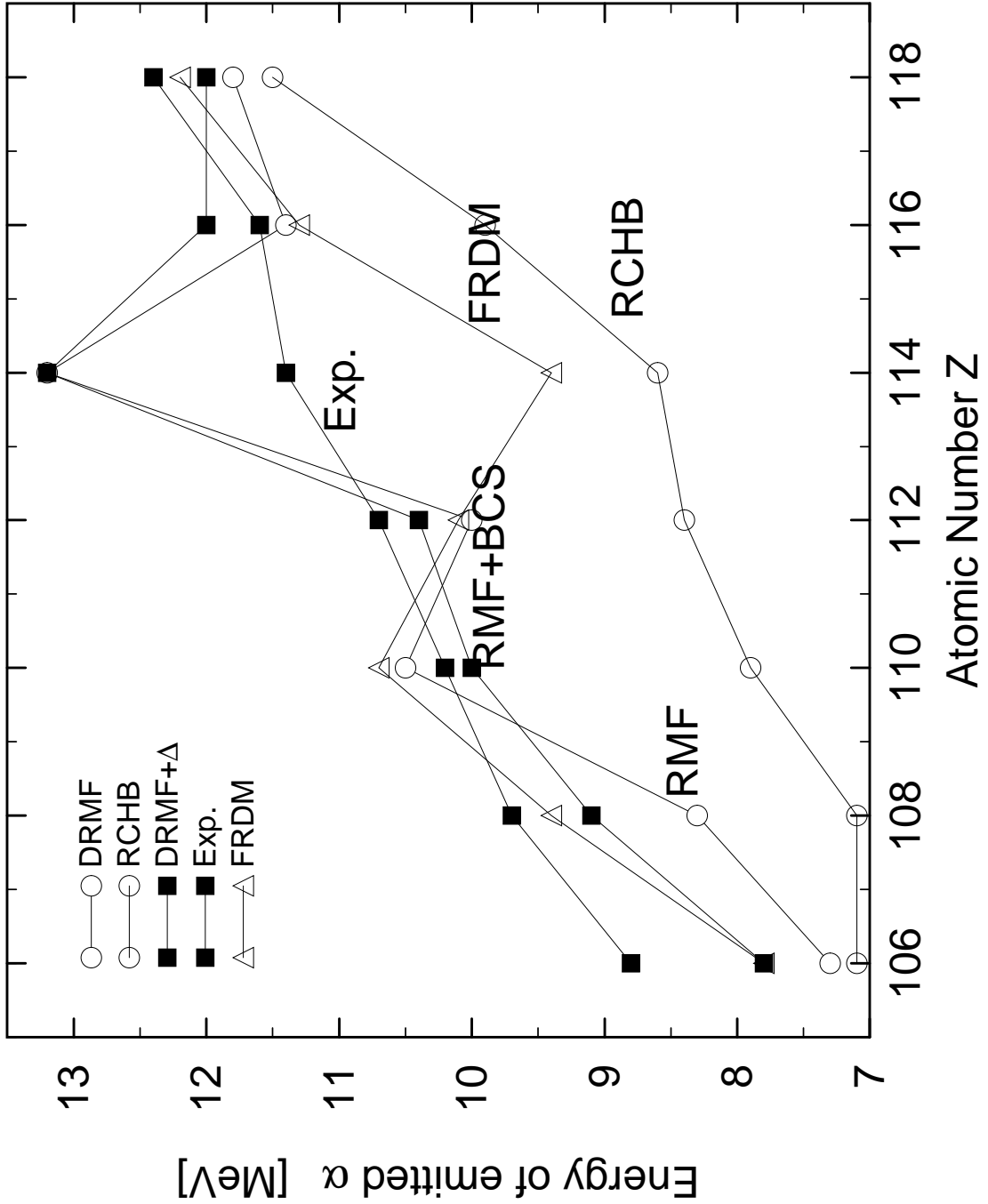}}}}
\vspace{8mm}
\caption{Comparison of the theoretical $\alpha$ particle energies 
$\Qa$ with the observed data.}
\label{QversusZ}
\end{figure}

The $\alpha-$decay energy $\Qa$ is 
shown in Fig.\ref{QversusZ} as a function of the atomic 
number along the decay chain from $^{293}$118.
The observed data and the prediction of FRDM are also included, 
where the former are taken from Fig.4 in \cite{Nino99}. 
Compared with the data, 
RCHB calculations give systematically too small $\Qa$. 
This reflects the deformation effect which is 
ignored in the present RCHB calculations. 
The $\Qa$ calculated in the RMF calculations
neglecting the pairing correlation but including the 
deformation (the open circles) is  somewhat larger than 
that in RCHB but still smaller  
than the data for $Z < 108$ and fluctuates for larger Z 
showing a sharp peak at $Z = 114$. 
This contrasts to the result of FRDM, which also fluctuates, 
but shows a deep minimum at the same atomic number 
reflecting a sub-closed shell at $Z = 114$ in this model.

\section{The description of superheavy elements by DRMF+BCS}

Using the pairing gap from RCHB, we now perform the RMF 
calculations by including both deformation 
and pairing correlation. 
The results are given in Table \ref{DRMF+BCS} for 
the binding energy $E$, the $\alpha-$particle energy, 
matter and charge radii, and the neutron, proton and matter 
deformation parameters.
The calculated binding energies for $^{86} Kr$, $^{208} Pb$
and  $^{294}118$ are also given.  
Each binding energy increases by 0.3 to 2 MeV with the pairing 
correlation, and can noticeably alter 
the atomic number dependence of $\Qa$.

We added the results of $\Qa$ calculated by the DRMF+BCS 
in Fig.\ref{QversusZ}. Comparing with RMF calculations, 
we observe that the theoretical $\Qa$ becomes much closer 
to the experimental data by the inclusion of the 
pairing correlation. 
Only for $Z = 114$, the $\Qa$ 
remains the same and has a difference of 2 MeV from the data.
Interestingly, $\Qa$ takes maximum at $Z = 114$ in DRMF+BCS 
in accord with the hump in the data, while the FRDM 
gives a minimum there.

\begin{table}
\caption{The binding energy $E$, $\alpha-$decay Q value, 
matter and charge radii, and neutron, proton and 
matter deformation parameters calculated in the DRMF+BCS theory with TM1.
}
\begin{center}
\label{DRMF+BCS}
\begin{tabular}{ccccccccccc}
$^A    Z   _N$    &&  $E$ & $\Qa$ & $R_m$ & $R_c$ & $\beta_n$ & $\beta_p$ & $\beta$ \\
\hline \hline
$^{ 86} Kr_{ 50}$ &&  751.0 &       & 4.222 & 4.189 & 0.003     & 0.005     &   0.004 \\
\hline
$^{208} Pb_{126}$ && 1636.8 &       & 5.650 & 5.544 & 0.000     & 0.0000    &   0.000 \\
\hline
$^{265}104_{161}$ && 1951.4 &       & 6.206 & 6.117 & 0.212     & 0.210     &   0.211 \\      
\hline
$^{269}106_{163}$ && 1971.9 & 7.8   & 6.232 & 6.147 & 0.208     & 0.207     &   0.208 \\
\hline
$^{273}108_{165}$ && 1991.1 & 9.1   & 6.256 & 6.175 & 0.197     & 0.197     &   0.197 \\
\hline
$^{277}110_{167}$ && 2009.4 & 10.0  & 6.279 & 6.202 & 0.178     & 0.178     &   0.178 \\
\hline
$^{281}112_{169}$ && 2027.3 & 10.4  & 6.302 & 6.228 & 0.164     & 0.164     &   0.164 \\
\hline
$^{285}114_{171}$ && 2042.4 & 13.2  & 6.347 & 6.280 & 0.204     & 0.202     &   0.203 \\
\hline
$^{289}116_{173}$ && 2058.7 & 12.0  & 6.390 & 6.328 & 0.227     & 0.228     &   0.227 \\
\hline
$^{293}118_{175}$ && 2075.0 & 12.0  & 6.423 & 6.363 & 0.230     & 0.234     &   0.232 \\
\hline
$^{294}118_{176}$ && 2081.3 &       & 6.429 & 6.365 & 0.224     & 0.229     &   0.226 \\
\end{tabular}
\end{center}
\end{table}


\section{Summary}

We made a systematic study of the 
structure of superheavy elements recently discovered
at Berkeley Lab's 88-Inch Cyclotron by the reaction 
$^{86}$Kr + $^{208}$Pb at 449 MeV 
in the framework of Relativistic Mean Field (RMF)
approach. We have shown that usually used
various RMF forces, which provide fair description of normal
stable nuclei, give quite different predictions for
superheavy elements. Among them TM1 is found to be the
good candidate to describe superheavy elements.

We have shown that the binding energy obtained from TM1 
agrees with that of FRDM within a difference of $2$ MeV.
The same conclusion that TM1 is the good interaction 
has been drawn from the calculations of the 
binding energy of Pb isotopes using Relativistic Continuum
Hartree Bogoliubov (RCHB) theory. 
However, neither the deformation nor the pairing correlation 
alone could explain the data of $\Qa$. 

We then performed 
RMF calculations of superheavy elements which 
include both the pairing correlation and deformation 
by using the pairing gaps obtained from RCHB.  
We have thus shown that 
a good agreement can be obtained between theory and experimental 
data concerning the Q value of the $\alpha$-decay. 
Especially, our RMF calculations reproduce a 
peak at Z=114 seen in the experimental data. 
We conjecture that this peak appears because 
of the shift of the shell structure, e.g. concerning nuclear shape, 
from Z=114 in FRDM to Z=112. 

Finally we wish to make a few comments on open questions.  
We kept the pairing gap parameter 
once it has been fixed for a spherical shape, 
ignoring the possibility of its 
shape dependence \cite{yt97}. Another basic assumption is that 
the observed $\alpha-$decays are from ground state to ground state, 
though this might not be the case for a part of the $\alpha-$decay chain. 
We noticed a paper by Cwiok et al. \cite{cwi99} 
after we have completed our study. 
The validity of the above mentioned approximation and assumption 
will be worth being examined 
to obtain more reliable understanding of superheavy elements. 
It would also be important to understand the difference 
between our conclusions and those in ref.\cite{cwi99}, 
where the authors predict systematically much smaller deformation 
for all nuclei and also claim that the deformation monotonically 
decreases towards Z=118. 
We will address these questions in a separate paper.

J.M. thank the Department of Physics, Tohouku university 
for its hospitality and the financial support of Japan Society for the 
Promotion of Science to make his stay possible.
This work was partially sponsored by the National Science Foundation
in China under Project No. 19847002.and by SRF for ROCS, SEM, China.


\end{document}